\documentclass[journal, a4paper, twoside]{IEEEtran}
\usepackage{amsmath,amssymb,amsfonts,amsthm}
\usepackage{algorithmic}
\usepackage{graphicx}
\usepackage{textcomp,multirow,longtable,pdflscape}
\usepackage{bbm}
\newtheorem{lemma}{\textbf{Lemma}}

\newtheorem{theorem}{\textbf{Theorem}}

\newtheorem*{prof}{\textbf{Proof}}
\newtheorem{Example}{\textbf{Example}}
\title{Private Information Delivery with Coded Storage}
\author{Kanishak Vaidya and B Sundar Rajan \\
Department of Electrical Communication Engineering, IISc Bangalore, India \\
E-mail: \{kanishakv, bsrajan\}@iisc.ac.in \vspace{-3mm}
}

\begin{document}
\maketitle
\thispagestyle{empty}

\begin{abstract}
     In private information delivery (PID) problem, there are $K$ messages stored across $N$ servers, each capable of storing $M$ messages and a user. Servers want to convey one of the $K$ messages to the user without revealing the identity (index) of the message conveyed. The capacity of PID problem is defined as maximum number of bits of the desired message that can be conveyed privately, per bit of total communication, to the user. For the restricted case of replicated systems, where coded messages or splitting one message into several servers is not allowed, the capacity of PID has been characterized by Hua Sun in ``Private Information Delivery, IEEE Transactions on Information Theory, December 2020" in terms of $K, N$ and $M.$ In this paper, we study the problem of PID with coded storage at the servers. For a class of problems called {\it bi-regular PID} we characterize the capacity for $N=K/M$ and for $N>K/M$ we provide an achievable scheme. In both the cases the rates achieved are more than the rates achievable with the replicated systems.  
\end{abstract}

\section{Introduction}
        The problem of Private Information Delivery (PID) was introduced in~\cite{Sun1,Sun2}. In PID, a dataset comprised of $K$ identically distributed messages is stored over $N$ servers. The servers want to convey one of the $K$ messages to a user but don't want the identity of message to be disclosed to the user. For example, the data stored at the servers could be medical records of patients from some hospitals. These hospitals want to send the record of one of the patients externally but want to preserve the privacy of the patient. 

    In order to convey a message to the user, without disclosing the identity of the message, servers may have to transmit more data than the actual size of the message. Because of this, the goal in a PID problem is to reduce \textit{transmission cost} while keeping the identity of the message private. The \textit{rate} of PID is defined as the ratio of the size of the message conveyed to the user to the amount of data sent to the user via transmissions to convey that message. Therefore in PID problems, the goal is to maximize the rate. 

    \subsection{Private Information Delivery~\cite{Sun2}}
        \begin{figure}[t]
            \begin{center}
                \includegraphics[width=0.4\textwidth]{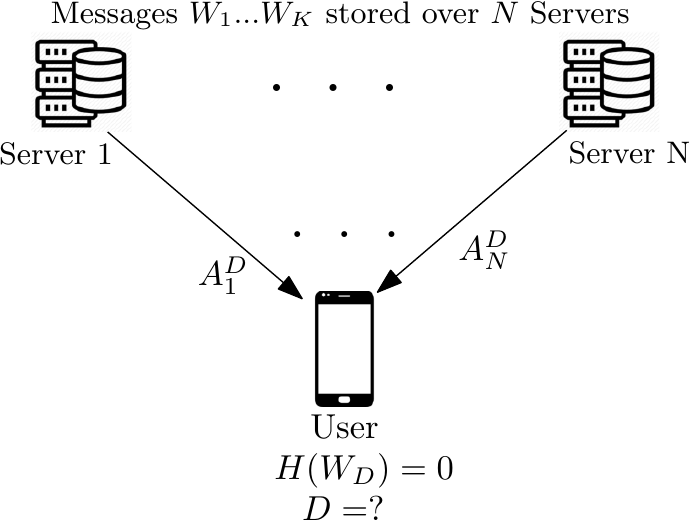}
                \caption{Private Information Delivery Problem}
            \end{center}
        \end{figure}
        
 In this subsection, a brief summary of the problem setup and results of the PID problem of~\cite{Sun2} is described. Consider a dataset comprised of $K$ independent messages $W_1 , W_2 \cdots W_K$, comprised of $L$ i.i.d.\ uniform symbols from the finite field $\mathbb{F}_q$ with $q$ elements for some integer $L$. There are $N$ servers, and each stores $M$ out of $K$ messages. The messages stored can not be coded messages and a message can not be split into sub-messages and stored in several servers. This is called a replicated system in \cite{Sun2}. Also, the servers share a common random variable $U$, which is independent of the messages. Servers privately generate an integer $D$ between $1$ and $K$ and wish to deliver $W_D$ to a user while keeping $D$ a secret from the user. In order to deliver $W_D$ to the user, server $n$ transmits $A_n^D$ to the user, which is a function of the messages stored at server $n$ and shared random variable $U$. The user will decode $W_D$ from all the $N$ transmissions it receives from the servers and should not be able to get any information about message index $D$. The rate for the PID scheme is defined as
        \[
            R \triangleq \frac{L}{\sum_{n=1}^N T_n}  
        \]
        where $T_n$ is the expected number of symbols sent from the server n to the user. The supremum of all achievable rates is called the capacity. For replicated systems, we denote the capacity by $C_{US}$ with the subscript standing for {\it Uncoded Storage}. The following results have been presented in \cite{Sun2}.

        \begin{theorem}\label{Sun1} ~\cite{Sun2}: For the PID problem with $K$ messages, $N \geq \lceil K/M \rceil$ servers and $M$ messages per server, the capacity satisfies
        \[
            1/\lceil K/M \rceil \leq C_{US} \leq M/K.
        \] 
\end{theorem}
        A converse of rate $M/K$ and an achievable scheme that achieves rate $ 1/\lceil K/M \rceil $ with $L = 1$ is provided in~\cite{Sun2}. For $K/M \in \mathbb{Z}$ Theorem \ref{Sun1} states that $C_{US} = M/K$. The next result is a condition on the number of servers such that the upper bound in Theorem \ref{Sun1}~\cite{Sun2} is tight

        \begin{theorem} \label{Sun2} ~\cite{Sun2}: For the PID problem with $K$ messages, $N \geq \lceil K/M \rceil$ servers and $M$ messages per server, and $K/M \notin \mathbb{Z}$, rate $M/K$ is achievable if
        \[
            N \geq \frac{K}{\gcd(K,M)} - \left( \frac{M}{\gcd(K,M) - 1} \right)\left(\left\lfloor \frac{K}{M} \right\rfloor - 1\right)
        \] 
        and rate $1 / \lceil K/M \rceil$ is optimal if $N = \lceil K/M \rceil$
\end{theorem}
        To prove Theorem \ref{Sun2} a PID scheme that achieve rate $M/K$ for $L = M / \gcd(K,M)$ and then for $N = \lceil K/M \rceil$ a proof of optimality of $R = 1 / \lceil K/M \rceil$ is given in~\cite{Sun2}. 
\subsection{Contributions}

        In the PID problems in \cite{Sun2} reviewed in the previous subsection whole messages are stored at the servers without coding or splitting. In this work, we consider the case of distributed systems, where a message can be accessed by a proper subset of servers and messages which are available to the servers can be split and coded before storing them at the servers. By using coded storage at the servers, we establish in the later sections  that the message can be delivered privately to the user while requiring less storage and transmitting less data. Specifically, for a class of problems called {\it bi-regular PID} we characterize the capacity for $N=K/M$ and for $N>K/M$ we provide an achievable scheme. In both the cases the rates achieved are more than the rates achievable with the replicated systems.
        
 \subsection{A motivating example}
 The following motivating example illustrates the advantages of keeping coded messages at the servers.
        
\begin{Example} 
Consider $K = 3$ messages uniformly distributed over $\mathbb{F}_5^2$, $N = 3$ servers, all privately sharing (i.e.\ without the user knowing about it) a random variable $u$ uniformly distributed over $\mathbb{F}_5$, and each server can store $2$ more symbols of $\mathbb{F}_5$, which means $M = 1$. Let $W_n = (W_{n,1} , W_{n,2}) , n \in \{1,2,3\}$ be the messages. We say that server $n$ is associated to message $W_k$ (and message $W_k$ is associated to server $n$) if server $n$ can store some information corresponding to $W_k$ in its storage. Let each message be associated to two servers. $W_1$ is associated to server 1 and server 2, $W_2$ is associated to server 2 and server 3 and $W_3$ is associated to server 3 and server 1. After storing two $ \mathbb{F}_5 $ symbols corresponding to two messages, servers will collectively choose a message index $D$ at random and wish to deliver $W_D$ to a user. The goal of the servers is to convey $W_D$ to the user privately and correctly i.e.\ the user should know the contents of the message but should not know the index of the message.

Now for this setup, we briefly describe a privacy preserving strategy that does not disclose any information about the index of the delivered message. Below we show a storage design in which parts of messages are not stored as such but first encoded as follows:
\begin{eqnarray*}
	C_{1,1} = 2W_{1,1} + 4W_{1,2} & C_{1,2} = 4W_{1,1} + W_{1,2} \\
	C_{2,1} = 3W_{2,1} + 4W_{2,2} & C_{2,2} = 3W_{2,1} + W_{2,2} \\
	C_{3,1} = 4W_{3,1} + 2W_{3,2} & C_{3,2} = 2W_{3,1} + 3W_{3,2}
\end{eqnarray*}
Following table enumerate the storage $Z_n$ at server $n$.
\begin{center}
	\begin{tabular}{|c|c|c|}
		\hline
		$Z_1$   &   $Z_2$   & $Z_3$ \\
		\hline
		$C_{1,1}$   &   $C_{1,2}$   &   $u$       \\
		$u$	    &	$C_{2,1}$   &	$C_{2,2}$   \\
		$C_{3,1}$   &	$u$	    &   $C_{3,2}$   \\
		\hline	
	\end{tabular}	
\end{center}
Each server is storing two symbols from $\mathbb{F}_5$ corresponding to two messages and a random $\mathbb{F}_5$ element. After storing these symbols, servers will collectively decide to transmit one of the message to the user. The table below enumerate what transmissions will be done by which server in order to convey message $W_D$ to the user where $A_1^D , A_2^D $ and $A_3^D$ are transmission from server $S_1 , S_2$ and $S_3$ respectively.  
\begin{center}
	\begin{tabular}{|c|c|c|c|}
		\hline
		For $D =$ &   $A_1^D$  &    $A_2^D$  &    $A_3^D$ \\
		\hline
		$1$ &   $C_{1,1} + u$    &   $C_{1,2,} + 3u$    &   $u$ \\
		$2$ &   $u$    &   $C_{2,1} + 3u$    &   $C_{2,2} + u$ \\
		$3$ &   $C_{3,1} + u$    &   $3u$    &   $C_{3,2} + u$ \\
		\hline
	\end{tabular}
\end{center}
Note that in order to convey one message, each server is transmitting one element of $\mathbb{F}_5$. 

Now to decode the message from these transmissions, the user will perform following computation:
\begin{align*}
	W_{D , 1} &= A_1^D + A_2^D + A_3^D \\
	W_{D , 2} &= A_1^D + 2A_2^D + 3A_3^D 
\end{align*}
It can be easily verified that the computation above can correctly convey any message to the user. It can also be easily seen that the message index $D$ is unknown to the user. That's because the user is receiving $3$ uniformly distributed symbols from $\mathbb{F}_5$ without any knowledge of to which message these symbols correspond to. And the user is performing the same computation, irrespective of the message index, to decode the message from the computation. Here we also note that decoding strategy is independent of message index $D$. And the user could not get information about the message index from the decoding strategy.

The rate achieved for the PID problem in this example is $\frac{1}{2}$ whereas the capacity $C_{US}$ is $\frac{1}{3}.$ 
\end{Example}

    \subsection{Fully Distributed System with message splitting}
        The problem of PID is trivial when every server can store a part of each message. As the minimum required memory is $M = \frac{K}{N}$ messages, the following scheme can achieve rate $1$ while using minimum possible memory. First, partition each message into $N$ sub-messages.
        \[
          W_{k} = \{W_{k , n} | n \in [N]\}
        \]
        Then server $n$ will store $W_{k , n} \forall n \in [N]$. As each server is storing $\frac{1}{N}$ fraction of each message, we have $M = \frac{K}{N}$. Then, if the servers choose to convey message $W_{\theta}$ to the user, they'll broadcast the part of the message they have in their storage. Specifically, server $n$ will send $W_{\theta , n}$. After receiving the transmissions of all the servers, the user will concatenate the transmissions to get the message, and it will not be able to get the index $\theta$.

        In this setup, the maximum possible rate, $R = 1$ is achieved, and minimum possible memory is used. This happens because each message is accessed by every server. In the rest of the paper, we consider cases where a message can't be accessed by every server. We also consider that the messages can be encoded before storing them at the servers.

    \subsubsection*{Notations}: For integers $a , b , c$ where $a \leq c , [a:b:c] \triangleq \{ a + nb : n \in \mathbb{Z}^+ , a + nb \leq c \}$. $[a:c]$ is same as $[a:1:c]$ and $[c]$ is same as $[1:c]$. For any $R \times C$ matrix $\textbf{M}$ and $\mathcal{N} \subseteq [C]$, $\textbf{M}_{\mathcal{N}}$ denotes the sub-matrix of $\textbf{M}$ formed using columns indexed by $\mathcal{N}$ and $ \textbf{M}_{r,c} $ denote element of $r^{th}$ row and $c^{th}$ column. For a set $\{A_1 , A_2 \hdots A_N\}$ indexed by integers between $1$ and $N$, $A_{ \mathcal{N} } , \mathcal{N} \subseteq [N] $ denotes subset $\{ A_n : n \in \mathcal{N}  \}$.
    
    
    \section{Problem Setup}
    Consider one user and $K$ independent messages $\{W_k\}_{k \in [K]}$. These messages are comprised of $L$ i.i.d.\ symbols from $\mathbb{F}_q.$ In the unit of $\mathbb{F}_q$ symbols, we have  
        
        \begin{align}\label{msg_length}
    	\begin{split}
    		H(W_k) &= L , \forall k \in [K] \\
    		H(\{W_k\}_{k \in [K]}) &= \sum_{k \in [K]} H(W_k) = KL
    	\end{split}
    \end{align}
    There are $N$ servers. We say that server $n$ is associated to message $W_k$ (and message $W_k$ is associated to server $n$) if server $n$ can store some information corresponding to $W_k$ in its storage. Every server can store up to $M$ messages. Let $Z_n$ denote the storage of server $n$. The storage of server $n$ depends on the messages associated to it and 
    \begin{equation}\label{storage}
    	H(Z_n) \leq ML , \forall n \in [N].
    \end{equation}
     Servers also store correlated random variables $U_n , n \in [N]$ which are independent of the messages. Server $n$ stores random variable $U_n.$ Let ${\bf U} =(U_1,U_2,\cdots,U_n).$ Then
    \begin{equation}\label{share_independence}
    	H({\bf U}, \{W_k\}_{k \in [K]}) = H({\bf U}) + H(\{W_k\}_{k \in [K]})
    \end{equation}
    Let, $\mathcal{N}_k$ denote the indices of the servers associated to message $W_k, k \in [K]$. We assume that every message is associated to $L \leq N$ servers, therefore $\mathcal{N}_k \subseteq [N] , |\mathcal{N}_k| = L$. 
    
    Let $\mathcal{K}_n $ denotes the set of messages, associated to the $n^{th}$ server where $ \mathcal{K}_n \subseteq [K]$. If every server is associated to same number of messages (i.e. $|\mathcal{K}_n| = |\mathcal{K}_1|, \forall n \in [N]$) then we call this problem to be $(K,N,M,L)$ bi-regular PID problem. Note that for bi-regular PID problems $KL/N \in \mathbb{Z}$ messages i.e. $|\mathcal{K}_n| = KL/N, \forall n \in [N]$. 
    
    Servers privately generate $D \in [K]$ and want to deliver $W_{D}$ to the user while keeping $D$ a secret from the user. In order to convey $W_{D}$ to the user, the $n^{th}$ server, $\forall n \in [N]$, will send transmission $A^{D}_n$ to the user. $A^{D}_n$ is completely determined from $Z_n$ and $U_n$ i.e.,
    \begin{equation}\label{AnswerConstruction}
    	H(A_{n}^D | Z_n , U_n , D) = 0.
    \end{equation}
    From $\{ A^{D}_n \}_{n \in [N]}$ the user has to decode $W_{D}$. So, it is required that
    \begin{equation}\label{correctness}
    	H(W_D | \{A^{D}_n\}_{n \in [N]}) = 0
    \end{equation}
    To decode $W_{D}$ without knowledge of $D$, the \textit{decoding strategy} employed by the user should be independent of $D$ (otherwise, decoding strategy will give information about $D$).  
    
    Now, in order to keep the desired index $D$ private from the user, it is required that 
    \begin{equation}\label{privacy}
    	I(D ; \{ A_n^D \}_{n \in [N]}) = 0
    \end{equation}  
    Performance of a PID is characterized by rate $R$ which is defined by the ratio of size of the message delivered to the user and total size of transmission performed by the serves. In our setting, the rate is 
    \begin{equation*}
    	R \triangleq \frac{L}{\sum_{n \in [N]} T_n}. 	
    \end{equation*}
    where $T_n$ is the size of transmission performed by server $n$ to the user i.e. $H(A_n^D) = T_n.$

    We also define $\eta_n \triangleq H(U_n)/L$ as the measure of size of randomness stored at the server $n$ for all $n \in [N]$. And a measure of size of total shared randomness $\textbf{U}$, is given by $\eta \triangleq H(\textbf{U})/L$. This shared randomness is not counted in server storage.
    
    Our goal is to construct \textit{PID scheme} with message splitting and coded messages that maximize $R$ while satisfying~\eqref{correctness} and~\eqref{privacy} under constraints~\eqref{msg_length},~\eqref{storage} and~\eqref{share_independence}. The maximum achievable rate with coded storage is called capacity and is denoted by $C_{CS}$.

    \begin{table}\label{parameter_comparison}
        \centering
        \caption{Restrictions on parameters of PID setup}
        \label{restrictiontable}
        \begin{tabular}{|c|c|}
            \hline
            Parameters in \cite{Sun2}   &   Parameters in our Setup \\  \hline \hline
            $N \in \mathbb{Z}$          &   $N \in \mathbb{Z}$ \\  \hline
            $K \in \mathbb{Z}$          &   $K \in \mathbb{Z}$ \\  \hline
            $M \in \mathbb{Z}$          & $M$ need not be an integer; \\
                                        &  $M \geq K/N$ \\ \hline
            $N \geq \lceil K/M \rceil$ &   $N \geq \lceil K/M \rceil$  \\
            \hline
              & \begin{tabular}{c}
                (For $L$-Biregular PID) \\   $L \in [\frac{N}{\gcd (N,K)}:\frac{N}{\gcd (N,K)}:N]$ \\
            \end{tabular} \\
            \hline
        \end{tabular}
    \end{table}

    \subsection*{Restrictions on PID Parameters}
        In $(K,N,M,L)$ bi-regular PID setup described above, we have already stated that $L \leq N$, and $KL/N \in \mathbb{Z}$. This will put a restriction on number of servers associated to a message that $L \in [\frac{N}{ \gcd (K,N) }: \frac{N}{ \gcd (K,N) } : N]$. Therefore if $K$ and $N$ are co-prime, then only possible value for $L$ is $N$. If $N$ divides $K,$ for example $K = 8$ and $N = 4,$ then $L$ can take all possible values less than $N$, i.e. $L \in [N]$. And as every server is associated to $|\mathcal{K}_n| = KL/N$ messages we get $|\mathcal{K}_n| \in [\frac{K}{ \gcd (K,N) }: \frac{K}{ \gcd (K,N) } : K]$. In Table \ref{restrictiontable} restrictions on PID parameters are given for our setting and the setting given in \cite{Sun2}. In Table \ref{L24x24} we have given possible values of $L$ for $K \in [2:24]$ and non prime values of $N \in [24]$.

        \begin{table*}
        \caption{Possible values of $L$ and $M \geq \frac{K}{N}$ for given $K$ and $N$}
        \label{L24x24}
        \centering                   
        \begin{tabular}{|c||c|c|c|c|c|c|c|c|c|c|c|c|c|c|}
        \hline
        K $\backslash$ N & 4 & 6 & 8 & 9 & 10 & 12 & 14 & 15 & 16 & 18 & 20 & 21 & 22 & 24 \\
        \hline
        \hline
        2    &   2,4 & 3,6 & 4,8 & 9      & 5,10 & 6,12 & 7,14 & 15      & 8,16 & 9,18 & 10,20 & 21      & 11,22 & 12,24 \\ \hline
        3    &   4      & 2,4,6 & 8      & 3,6,9 & 10      & 4,8,12 & 14      & [5:5:15] & 16      & 6,12,18 & 20      & 7,14,21 & 22      & [8:8:24] \\  \hline
        4    &   [4]     & 3,6 & [2:2:8] & 9      & 5,10 & [3:3:12] & 7,14 & 15      & [4:4:16] & 9,18 & [5:5:20] & 21      & 11,22 & [6:6:24] \\ \hline
        5    &   4      & 6      & 8      & 9      & [2:2:10] & 12      & 14      & [3:3:15] & 16      & 18      & [4:4:20] & 21      & 22      & 24  \\  \hline
        6    &   2,4 & [6]     & 4,8 & 3,6,9 & 5,10 & [2:2:12] & 7,14 & [5:5:15] & 8,16 & [3:3:18] & 10,20 & 7,14,21 & 11,22 & [4:4:24] \\ \hline
        7    &   4      & 6      & 8      & 9      & 10      & 12      & [2:2:14] & 15      & 16      & 18      & 20      & [3:3:21] & 22    & 24  \\  \hline
        8    &   [4]     & 3,6 & [8]     & 9      & 5,10 & [3:3:12] & 7,14 & 15      & [2:2:16] & 9,18 & [5:5:20] & 21      & 11,22 & [3:3:24] \\  \hline
        9    &   4      & 2,4,6 & 8      & [9]     & 10      & 4,8,12 & 14      & [5:5:15] & 16      & [2:2:18] & 20      & 7,14,21 & 22      & [8:8:24] \\  \hline
        10   &   2,4 & 3,6 & 4,8 & 9      & [10]     & 6,12 & 7,14 & [3:3:15] & 8,16 & 9,18 & [2:2:20] & 21      & 11,22 & 12,24 \\  \hline
        11   &   4      & 6      & 8      & 9      & 10      & 12      & 14      & 15      & 16      & 18      & 20      & 21      & [2:2:22] & 24  \\  \hline
        12   &   [4]     & [6]     & [2:2:8] & 3,6,9 & 5,10 & [12]     & 7,14 & [5:5:15] & [4:4:16] & [3:3:18] & [5:5:20] & 7,14,21 & 11,22 & [2:2:24] \\  \hline
        13   &   4      & 6      & 8      & 9      & 10      & 12      & 14      & 15      & 16      & 18      & 20      & 21      & 22      & 24  \\ \hline
        14   &   2,4 & 3,6 & 4,8 & 9      & 5,10 & 6,12 & [14]     & 15      & 8,16 & 9,18 & 10,20 & [3:3:21] & 11,22 & 12,24 \\  \hline
        15   &   4      & 2,4,6 & 8      & 3,6,9 & [2:2:10] & 4,8,12 & 14      & [15]     & 16      & 6,12,18 & [4:4:20] & 7,14,21 & 22      & [8:8:24] \\  \hline
        16   &   [4]     & 3,6 & [8]     & 9      & 5,10 & [3:3:12] & 7,14 & 15      & [16]     & 9,18 & [5:5:20] & 21      & 11,22 & [3:3:24] \\  \hline
        17   &   4      & 6      & 8      & 9      & 10      & 12      & 14      & 15      & 16      & 18      & 20      & 21      & 22      & 24  \\ \hline
        18   &   2,4 & [6]     & 4,8 & [9]     & 5,10 & [2:2:12] & 7,14 & [5:5:15] & 8,16 & [18]     & 10,20 & 7,14,21 & 11,22 & [4:4:24] \\   \hline
        19   &   4      & 6      & 8      & 9      & 10      & 12      & 14      & 15      & 16      & 18      & 20      & 21      & 22      & 24 \\ \hline
        20   &   [4]     & 3,6 & [2:2:8] & 9      & [10]     & [3:3:12] & 7,14 & [3:3:15] & [4:4:16] & 9,18 & [20]     & 21      & 11,22 & [6:6:24] \\ \hline
        21   &   4      & 2,4,6 & 8      & 3,6,9 & 10      & 4,8,12 & [2:2:14] & [5:5:15] & 16      & 6,12,18 & 20      & [21]     & 22      & [8:8:24] \\ \hline
        22   &   2,4 & 3,6 & 4,8 & 9      & 5,10 & 6,12 & 7,14 & 15      & 8,16 & 9,18 & 10,20 & 21      & [22]     & 12,24 \\  \hline
        23   &   4      & 6      & 8      & 9      & 10      & 12      & 14      & 15      & 16      & 18      & 20      & 21      & 22      & 24      \\ \hline
        24   &   [4]     & [6]     & [8]     & 3,6,9 & 5,10 & [12]     & 7,14 & [5:5:15] & [2:2:16] & [3:3:18] & [5:5:20] & 7,14,21 & 11,22 & [24]     \\  \hline
        \end{tabular}
        \end{table*}

\section{Main Results}
    In this section we state the optimal rate for $(K , N , M , L)$ bi-regular PID setup for $M = \frac{K}{N}$. And for general $K , N , M$ and $L \leq N$ such that $KL/\lceil K/M \rceil \in \mathbb{Z}$, we give an achievable rate. 

    For the optimal rate, we are taking $M = K/N$, which is the minimum required memory to store all the messages. As in bi-regular PID setup, $KL/N$ messages are associated to each server we have $KL/N = ML \in \mathbb{Z}$.

    \begin{theorem}\label{capacity}
        For $(K , N , M , L)$ bi-regular PID setup with $M = K/N$, the PID capacity is given by
        \begin{equation*}
            C_{CS} = \frac{ML}{K}
        \end{equation*} \hfill \qedsymbol
    \end{theorem}

    Note that if $ML = K$ then  $\frac{KL}{N} = K$ or $L = N$. This means, each message is associated to all the servers, which is a fully distributed system, which can achieve rate $1$ as discussed earlier. Also note that as $M = \frac{K}{N}$ the capacity is also given by $C_{CS} = \frac{L}{N}$

    To prove Theorem~\ref{capacity} we first give an achievable scheme (in Section~\ref{scheme}) that achieve rate $L/N$ and then give an information theoretic proof (Section~\ref{converse}) that rate higher than $L/N$ cannot be achieved.

    Theorem \ref{capacity} gives PID capacity for bi-regular PID setup if $N = K/M$. But for the case when $N > K/M$, we give an achievable rate for PID setup with coded storage 

    \begin{theorem}\label{rate}
        For PID setup with $K$ messages, $N$ servers, $M$ messages per server, $N > K/M$ and each message associated to $L$ servers with $KL / \lceil K/M \rceil \in \mathbb{Z}$, the following PID rate can be achieved
        \begin{equation*}
            R = \begin{cases}
                \frac{L}{\lceil K / M \rceil} & \mbox{if } L < \lceil K/M \rceil \\ 
                1 & \mbox{if } L \geq \lceil K/M \rceil
            \end{cases}
        \end{equation*}
    \end{theorem}

    \begin{prof}
        As $N > \frac{K}{M}$, consider any $\lceil K/M \rceil$ servers. If $L < \lceil K/M \rceil$ then there exist a $(K , \lceil K/M \rceil , \frac{K}{\lceil K/M \rceil} , L)$ bi-regular PID setup corresponding to $\lceil K/M \rceil$ servers and $K$ messages that can achieve the rate 
        \[
           R = \frac{K}{\lceil K/M \rceil} \frac{L}{K} = \frac{L}{\lceil K/M \rceil} 
        \] as described in Theorem \ref{capacity}. In this case, any of the $\lceil K/M \rceil$ servers will store $\frac{K}{\lceil K/M \rceil}$ message each and only these $\lceil K/M \rceil$ servers transmit to the user, while remaining $N - \lceil K/M \rceil$ servers won't transmit. 

        If $L \geq \lceil K/M \rceil$, then there exists a fully distributed system, corresponding to a subset of $\lceil K/M \rceil$ servers, which are capable of storing all the messages, and hence the PID rate $R = 1$ is achievable. \hfill \qedsymbol
    \end{prof}

    \textbf{Remark:} If $K/M \in \mathbb{Z}$, then in PID setup with $K$ messages, $N$ servers where $N \geq K/M$, $M$ messages per server and $L$ servers associated to each message such that $ML \in \mathbb{Z}$, the rate $R = ML/K$ can be achieved if $L < K/M$ and $R = 1$ is achievable if $L \geq K/M$. And this rate is optimal if $N = K/M$.

    Furthermore, in the achievable scheme (in Section~\ref{scheme}), capacity is achieved with $\eta = \frac{K}{ML} - 1$ and $\eta_n = \frac{1}{L} , \forall n \in [N]$.

\section{Comparison with Previous Work}
    In this section we  compare our results with those in ~\cite{Sun2}.

\begin{figure}
        \centering
        \includegraphics[width=0.4\textwidth]{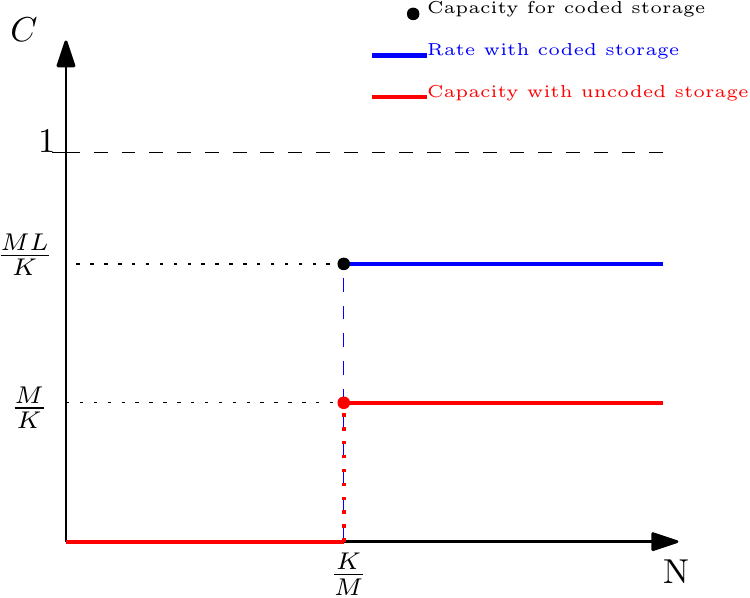}
        \caption{Capacity vs number of servers plot for given $K , L , M$ where $K/M \in \mathbb{Z}$ and $ML \in \mathbb{Z}$. We are comparing our results with PID setup with uncoded storage \cite{Sun2} for which (red curve) capacity is fully characterized. We have achieved capacity for point $N = K/M$ (black) in bi-regular PID setup with coded storage. Capacity is $C_{CS} = ML/K$ which is $L$ times larger than the capacity achieved in \cite{Sun2} for the same memory point. The PID capacity with coded storage for $N > K/M$ (blue) is not known but rate achieved is $L$ times higher than the capacity of PID setup with uncoded storage.}
        \label{CvsN}
    \end{figure}

 Theorem 1 of~\cite{Sun2} states that for $N \geq \lceil \frac{K}{M} \rceil $, the capacity $C_{US}$ satisfies
    \[
        1/\lceil K/M \rceil \leq C_{US} \leq M/K.  
    \]
    To compare our scheme with the scheme given in~\cite{Sun2}, consider a setup where $N = K/M \in \mathbb{Z}$. For this setup, as seen in Figure \ref{CvsN}, with uncoded storage, the capacity is 
    \[
        C_{US} = \frac{M}{K}. 
    \]
    But if we allow splitting and encoding of messages, i.e.\ in our setting, where $M = K/N$ messages are stored at the server and $L$ servers are associated to each message, the PID capacity
    \[
        C_{CS} = \frac{ML}{K}  
    \]
    is achieved. 

    When $N > K/M$ we see that, with coded storage, again, rate achieved is $L$ times larger than the capacity of PID setup with uncoded storage.

    \begin{figure}
        \centering
        \includegraphics[width=0.4\textwidth]{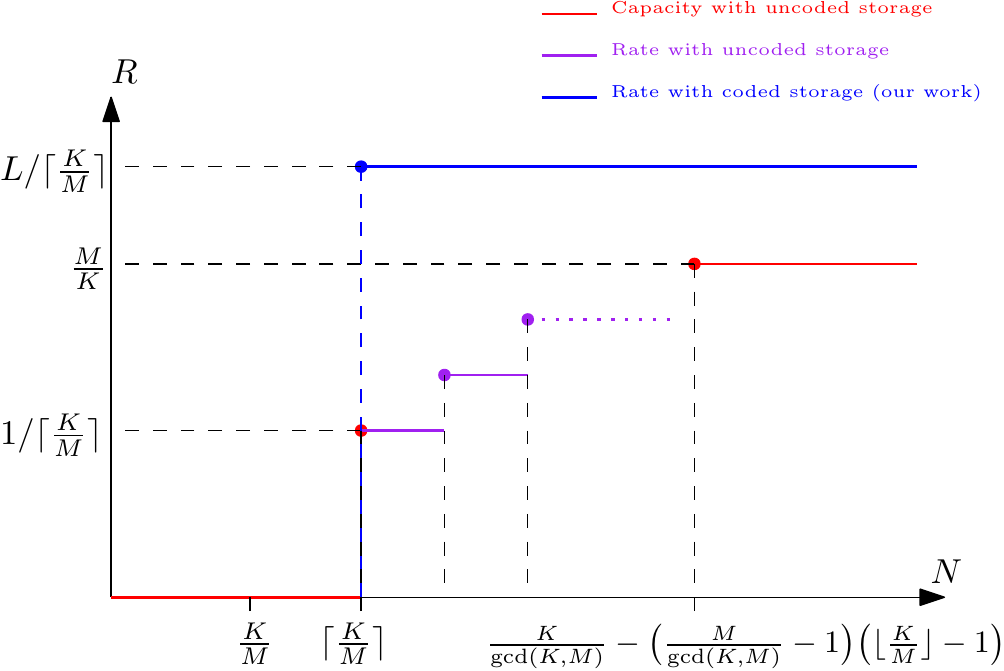}
        \caption{Rate $R$ vs number of servers plot for given $K , L , M$ where $K/M \not \in \mathbb{Z}$ and $\frac{KL}{\lceil K/M \rceil} \in \mathbb{Z}$. For PID setup with uncoded storage \cite{Sun2} (red curve) capacity is $C_{US} = \frac{M}{K}$ if $N \geq \frac{K}{\gcd (K,M)} - \big( \frac{M}{\gcd (K,M)} - 1\big) \big( \lfloor \frac{K}{M} - 1 \rfloor \big)$ and $C_{US} = 1 / \lceil K / M \rceil$ if $N = \lceil K / M \rceil$. For smaller number of servers and $N > \lceil \frac{K}{M} \rceil$ rate between $M/K$ and $1 / \lceil K/M \rceil$ is achievable. We are achieving rate $L/\lceil K/M \rceil$ when $N \geq \lceil K/M \rceil$.}
        \label{RvsN}
    \end{figure}

    When $K/M \not \in \mathbb{Z}$ and $\frac{KL}{\lceil K/M \rceil} \in \mathbb{Z}$, as shown in Figure \ref{RvsN}, we are achieving rate $R = \frac{L}{\lceil K/M \rceil}$ when $N \geq \lceil \frac{K}{M} \rceil$ with coded storage. With uncoded storage \cite{Sun2} maximum achievable rate is $M/K$ which is less than $\frac{L}{\lceil K/M \rceil}$ for $L \geq 2$. Furthermore if $L \geq \lceil K/M \rceil$ then rate $R = 1$ is achievable.

    Although for rate calculation, we are only considering server storage as number of messages stored in a server i.e.\ $M$, but servers are also storing correlated random variables $U_1 \hdots U_N$. In achievable scheme given in~\cite{Sun1}, rate $M/K$ is achieved when $N = \frac{K}{\gcd(K,M)} - \big(\frac{M}{\gcd(K,M)} - 1\big)\big(\lfloor \frac{K}{M} - 1 \rfloor \big)$ with $\eta = 1/R - 1 = K/M - 1$. Whereas, in achievable scheme presented in Section~\ref{scheme}, we have $\eta = H(\textbf{U}) / L = K/ML - 1$ and $\eta_n = 1/L, \forall n \in [N]$. 

    For example, consider the case where $N = 6$ servers, $K = 12$ messages, and each server can store $M = 2$ messages and $L = 4$. Then if we choose to store two messages per server without encoding or splitting, we can achieve rate $2/12 = 1/6$, but if encoding and splitting are allowed, rate $8/12 = 4/6$ can be achieved while storing the same number of messages at each server. Also, the shared randomness required in the uncoded storage case is five times the size of one individual message. In contrast, with coded storage, achievable scheme only requires shared randomness, which is half the size of an individual message. Furthermore, in coded storage scheme, each server will store a random variable, which is $1/4^{th}$ of an individual message in size.

    Now consider the example (explained in details in Section~\ref{schemeexample}) with $N = 6$ servers and $K = 8$ messages. With coded storage and $L=3$ servers associated to each message, the PID rate $0.5$ can be achieved with $M = 4/3$ messages stored per server, while with uncoded storage \cite{Sun2} the PID capacity is $0.375$ which is achieved while storing $M = 3$ messages in every server. Thus, by splitting and coding messages before storing them on the server, we achieve a higher rate while using lesser memory than the scheme that uses uncoded storage at the servers. Also, the scheme with coded storage requires $\eta = 1$, which means the size of shared randomness is equal to the size of one message, and $\eta_n = 1/3, \forall n \in [N]$ which means every server have to store extra one-third of a message as randomness besides $4/3$ messages they are already storing. Whereas with the uncoded storage scheme with $M=3$, randomness shared amongst servers is $5/3$ times more than the size of a single message.

\section{Achievable Scheme}\label{scheme}
    In this section, we present an achievable scheme for the $(K,N,K/N,L)$ bi-regular PID problem achieving rate $L / N$. First we illustrate the scheme with an example.
    \subsection{Example}\label{schemeexample}
        Consider $K = 8$ messages $\{W_k\}_{k \in [8]}$, $N = 6$ servers $\{S_n\}_{n \in [6]}$ and $L = 3$ associated servers per message. Let, each message consist of $3$ symbols of $ \mathbb{Z}_{11} $ (integers modulo 11). In this setup, each server can store $M = K / N = 4 / 3$ messages i.e. $4$ elements of $ \mathbb{Z}_{11}$. Let the messages be: 
        \begin{equation*}
            W_k = \begin{bmatrix}
                W_{k,1} \\
                W_{k,2} \\
                W_{k,3}
            \end{bmatrix}
            , \forall k \in [8].
        \end{equation*}
        Every message is associated to $3$ servers and every server is associated to $4$ messages. In the table below we list the indices of the servers that are associated to a message i.e. $ \mathcal{N}_k$.
        
        \begin{center}
            \begin{tabular}{|c|c|}
                \hline
                Message $W_k$  &   Indices of the associated servers $ \mathcal{N}_k $ \\
                \hline
                $W_1$  &   $\{ 1 , 2 , 3\}$ \\
                $W_2$  &   $\{ 1 , 2 , 3\}$ \\
                $W_3$  &   $\{ 1 , 2 , 3\}$ \\
                $W_4$  &   $\{ 1 , 2 , 3\}$ \\
                $W_5$  &   $\{ 4 , 5 , 6\}$ \\
                $W_6$  &   $\{ 4 , 5 , 6\}$ \\
                $W_7$  &   $\{ 4 , 5 , 6\}$ \\
                $W_8$  &   $\{ 4 , 5 , 6\}$ \\
                \hline
            \end{tabular}
        \end{center}
        Now consider the MDS code having parity-check matrix
        \begin{equation*}
            \textbf{H} = \begin{bmatrix}
                1   &   1   &   1   &   1   &   1   &   1   \\
                1   &   2   &   3   &   4   &   5   &   6   \\
                1   &   4   &   9   &   5   &   3   &   3   \\
            \end{bmatrix}
        \end{equation*}
        and the generator matrix
        \begin{equation*}
            \textbf{G} = \begin{bmatrix}
                3  &  8  &  1  &  7  &  2  &  1 \\
                3  &  4  &  4  &  0  &  1  & 10 \\
                6  & 10  &  6  &  5  &  1  &  5 \\
            \end{bmatrix}
        \end{equation*}
        both matrices over $ \mathbb{Z}_{11}$. 
        
        Servers will also store a variable as follows: Consider a random variable $\textbf{U} = {(u_1 , u_2, u_3)}^{\top}$, generated privately (i.e. without the user knowing about it) where the entries are uniformly chosen from $\mathbb{Z}_{11}$. Let $U_n = g_n^{\top}\textbf{U}$ where $g_n$ is the $n^{th}$ column of $\textbf{G}$. Then server $n$ will store $U_n$. 
        
        Message $W_k$ will be encoded as $C_k$ where 
        \[ C_k = \textbf{H}_{ \mathcal{N}_k}^{-1} W_k \] 
        i.e. 
        \[
            C_k = \begin{bmatrix}
                1   &   1   &   1   \\ 
                1   &   2   &   3   \\
                1   &   4   &   9   \\ 
            \end{bmatrix}^{-1}
            \begin{bmatrix}
                W_{k,1} \\ W_{k,2} \\ W_{k,3}
            \end{bmatrix} = \begin{bmatrix}
                3W_{k,1}    +   3W_{k,2}    -   5W_{k,3} \\
               -3W_{k,1}    +   4W_{k,2}    -    W_{k,3} \\
                 W_{k,1}    +   4W_{k,2}    -   5W_{k,3} \\
            \end{bmatrix}
        \]
        for $k \in \{1 , 2 , 3\}$ and 
        \[
            C_k = \begin{bmatrix}
                1   &   1   &   1   \\ 
                4   &   5   &   6   \\
                5   &   3   &   3   \\ 
            \end{bmatrix}^{-1}
            \begin{bmatrix}
                W_{k,1} \\ W_{k,2} \\ W_{k,3}
            \end{bmatrix} = \begin{bmatrix}
                4W_{k,1}    +   0W_{k,2}    -   5W_{k,3} \\
               -2W_{k,1}    -    W_{k,2}    -    W_{k,3} \\
                -W_{k,1}    +    W_{k,2}    -   5W_{k,3} \\
            \end{bmatrix}
        \]
        for $k \in \{4 , 5 , 6\}$. Now each server will store one symbol corresponding to each message it is associated to. Every server will store four $ \mathbb{Z}_{11} $ symbols. The following table enumerate storage of each server.
        \begin{center}
            \begin{tabular}{|c|c|}
                \hline
                Server      &   Storage $Z_n$\\
                \hline
                Server $1$  &   $\{ C_{k,1} : k \in [1:4]\}$ \\
                Server $2$  &   $\{ C_{k,2} : k \in [1:4]\}$ \\
                Server $3$  &   $\{ C_{k,3} : k \in [1:4]\}$ \\
                Server $4$  &   $\{ C_{k,1} : k \in [5:8]\}$ \\
                Server $5$  &   $\{ C_{k,2} : k \in [5:8]\}$ \\
                Server $6$  &   $\{ C_{k,3} : k \in [5:8]\}$ \\
                \hline
            \end{tabular}
        \end{center}
   After storing these messages, servers choose $D \in [8]$ uniformly. If $D = 1$ then the following transmissions are made by the servers in order to deliver $W_1$ to the user privately.
        \begin{center}
            \begin{tabular}{|c|c|}
                \hline
                Server n  &   Transmission $A_n^1$\\
                \hline
                Server $1$  &   $C_{1,1} + 3u_1 + 3u_2 - 5u_3$ \\
                Server $2$  &   $C_{1,2} + 8u_1 + 4u_2 -  u_3$ \\
                Server $3$  &   $C_{1,3} +  u_1 + 4u_2 - 5u_3$ \\
                Server $4$  &   $7u_1 + 0u_2 + 5u_3$ \\
                Server $5$  &   $2_1 +  u_2 +  u_3$ \\
                Server $6$  &   $ u_1 -  u_2 + 5u_3$ \\
                \hline
            \end{tabular}
        \end{center}
        Let $ \textbf{A}^D = [A_1^D , A_2^D \hdots A_6^D]^{\top}$ denote the transmission vector where $A_n^D$ is the transmission by the server $n.$ Then, for $D=1$ we have $ \textbf{A}^1 = [A_1^1 , A_2^1 \hdots A_6^1]^{\top}$ and the user will perform the following computation:
        \begin{align*}
            \textbf{HA}^1 &=  \begin{bmatrix}
                1   &   1   &   1   &   1   &   1   &   1   \\
                1   &   2   &   3   &   4   &   5   &   6   \\
                1   &   4   &   9   &   5   &   3   &   3   \\
            \end{bmatrix} 
            \begin{bmatrix}
               C_{1,1} + 3u_1 + 3u_2 - 5u_3 \\
               C_{1,2} + 8u_1 + 4u_2 -  u_3 \\
               C_{1,3} +  u_1 + 4u_2 - 5u_3 \\
                  7u_1 + 0u_2 + 5u_3 \\
                  2u_1 +  u_2 +  u_3 \\
                   u_1 -  u_2 + 5u_3 \\
            \end{bmatrix} \\
            &= \begin{bmatrix}
                C_{1,1} +  C_{1,2} +  C_{1,3} \\ 
                C_{1,1} + 2C_{1,2} + 3C_{1,3} \\ 
                C_{1,1} + 4C_{1,2} - 2C_{1,3} \\ 
            \end{bmatrix} = \begin{bmatrix}
                W_{1,1} \\
                W_{1,2} \\
                W_{1,3} \\
            \end{bmatrix}
        \end{align*}

        Note that every server is transmitting only one symbol of $ \mathbb{Z}_{11}$. Therefore the rate achieved in this example is $3 / 6 = 0.5$. 

\subsection{General Description}
Let the message $W_k \in \mathbb{F}_q^L$ be associated to servers indexed by $ \mathcal{N}_k $ where $ \mathcal{N}_k \subseteq [N] , | \mathcal{N}_k | = L$. Let such  associations be given by the invertible mappings $f_k : [L] \rightarrow \mathcal{N}_k , \forall k \in [K]$, for every message such that $\{ f_k(l) : l \in [L] \} = \mathcal{N}_k$. That is, $f_k(l)$ will be an index of a server associated to mesage $W_k$. This map is reversible, and for every $f_k$ there exist a reverse map $f_k^{-} : \mathcal{N}_k \rightarrow [L]$ such that $f_k^{-} (f_k(l)) = l$.

Now consider an MDS code $ \mathbb{C} $ having the parity check matrix $ \textbf{H} \in \mathbb{F}_q^{L \times N} $ and the generator matrix $ \textbf{G} \in \mathbb{F}_q^{N-L \times N} $. The sub-matrix $ \textbf{H}_{ \mathcal{N}_k } \in \mathbb{F}_q^{L \times L} $ formed by the columns of $ \textbf{H} $ indexed by $ \mathcal{N}_k $ will be invertible for all $k$. 

Server $n$ stores a random variable $u_n \in \mathbb{F}_q$ as follows: Choose a vector $\textbf{U}$ from $\mathbb{F}_q^{N-L}$ uniformly and randomly. Then server $n$ will store $U_n = g_n^{\top}\textbf{U}$, where $g_n$ is the $n^{th}$ column of $ \textbf{G} $. 

The message $W_k$ is encoded to $C_k \in \mathbb{F}_q^{L}  $ given by
\begin{equation*}
    C_k = \textbf{H}_{ \mathcal{N}_k }^{ -1 } W_k
\end{equation*}
and then the $l^{th}$ symbol of $C_k$ i.e. $C_{k,l}$ is stored at server $f_k(l)$. In $(K,N,K/N,L)$ bi-regular PID setup, every server is associated to $KL/N$ messages, and hence each server will be storing $KL/N$ symbols from $\mathbb{F}_q$, which is equivalent of storing $M = K/N$ messages.

Now, the servers collectively choose $D \in [K]$ and wish to convey $W_{D}$ to the user while keeping $D$ secret from the user. For that, $S_n$ will transmit $A_n^{D}$ to the user, where
\begin{align*}
    A_n^{D} &= C_{D , f^{-}_D(n)} + U_n   ~~~ \text{if} ~~~ n \in \mathcal{N}_{D}, \\
    A_n^{D} &= U_n ~~~ \text{if}~~~  n \in [N] \setminus \mathcal{N}_{D}.
\end{align*}
Let
\begin{equation*}
	\mathbf{A}^{D} \triangleq \begin{bmatrix}
	A^{D}_1 \\
	A^{D}_2 \\
	\vdots \\
	A^{D}_N \\
	\end{bmatrix}.
\end{equation*}

Now, the user will be able to decode $W_{D}$ from $\mathbf{A}^{D}$ as shown below.
\subsection*{Proof of Correctness}
The user will perform the computation: $\mathbf{H} \mathbf{A}^D.$

\textbf{Claim}:  $	\mathbf{H} \mathbf{A}^{D} = W_{D}.$ 

\begin{prof}  Letting $\textbf{h}_n$ denote the $n^{th}$ column of $\textbf{H},$ we have
    \begin{align*}
        \mathbf{H} \mathbf{A}^{D} &= \sum_{n =1}^{N} \mathbf{h}_n A_{n}^{D} \\
        &= \sum_{n \in \mathcal{N}_{D}}\mathbf{h}_n A_{n}^{D} + \sum_{n \in [N]\setminus\mathcal{N}_{D}} \mathbf{h}_n A_{n}^{D} \\
        &= \mathbf{H}_{\mathcal{N}_{D}} A_{\mathcal{N}_{D}}^{D} + \sum_{n \in [N]\setminus\mathcal{N}_{D}} \mathbf{h}_n A_{n}^{D} \\
        &= \mathbf{H}_{\mathcal{N}_{D}} (\mathbf{C}_{D} + \mathbf{G}_{\mathcal{N}_{D}}^{\top}\textbf{U}) + \mathbf{H}_{[N] \setminus \mathcal{N}_{D}} \mathbf{G}_{[N] \setminus \mathcal{N}_{D}}^{\top}\textbf{U} \\
        &= \mathbf{H}_{\mathcal{N}_{D}} \mathbf{H}_{\mathcal{N}_{D}}^{-1} W_{D} \\
        &+ (\mathbf{H}_{\mathcal{N}_{D}} \mathbf{G}_{\mathcal{N}_{D}}^{\top} + \mathbf{H}_{[N] \setminus \mathcal{N}_{D}} \mathbf{G}_{[N] \setminus \mathcal{N}_{D}}^{\top})\textbf{U} \\
        &= W_{D}.
    \end{align*}  \hfill \qedsymbol
\end{prof}

\subsection*{Proof of Privacy}
Now we show that our scheme satisfies the privacy constraint given in \eqref{privacy}. 
\begin{prof}
    We have $I(D ; \textbf{A}^D ) = H(\textbf{A}^D) - H(\textbf{A}^D | D).$   We proceed to show that regardless of the value of $D$, $ \textbf{A}^D $ is uniformly distributed over all possible values (i.e. over $ \mathbb{F}_q^N $).
    
    For some $ \textbf{a} \in \mathbb{F}_q^N $ and $d \in [K],$  we have
    \begin{alignat*}{2}
        &\mathbb{P} \{ \textbf{A}^D &&= \textbf{a} | D = d\} \\
        &= \mathbb{P}\{ &&G_{ \mathcal{N}_D }^{\top}\textbf{U} + C_D = \textbf{a}_{\mathcal{N}_D} , G_{[N] \setminus \mathcal{N}_D }^{\top}\textbf{U} = \textbf{a}_{[N] \setminus \mathcal{N}_D} | D = d \} \\
        &= \mathbb{P}\{ &&W_D = H_{\mathcal{N}_D}(\textbf{a}_{\mathcal{N}_D} - G_{ \mathcal{N}_D }^{\top} G_{[N] \setminus \mathcal{N}_D }^{-\top} \textbf{a}_{[N] \setminus \mathcal{N}_D} ), \\
        & &&\textbf{U} = G_{[N] \setminus \mathcal{N}_D }^{-\top} \textbf{a}_{[N] \setminus \mathcal{N}_D} | D = d\}.
    \end{alignat*}
    As the message $W_D$, shared randomness $\textbf{U}$ and the message index $D$ are all mutually independent, we get 
    $$ \mathbb{P} \{ \textbf{A}^D = \textbf{a} | D = d\} = \frac{1}{q^N} , \forall d \in [K].$$
This implies $$H( \textbf{A}^D | D ) = \mathbb{E}\{ H(\textbf{A}_n^d) \} = N.$$ 
Since $N \geq H(\textbf{A}^D) \geq H(\textbf{A}^D | D) = N,$ we conclude that  $I(D; \textbf{A}^D ) = 0.$ \hfill \qedsymbol
\end{prof}

\textbf{Note}: Observe that the achievable scheme is independent of the fact that every server is associated to $KL/N$ messages and only require that each message is associated to $L$ servers. Therefore, given scheme will also work if servers don't have any restrictions on the number of messages they have access to as long as servers have enough storage to store the coded symbols, corresponding to the messages they are associated with. 

Also note that $\eta = \frac{H(\textbf{U})}{L} = \frac{N - L}{L} = \frac{K}{ML} - 1$ for a $(K,N,M,L)$ bi-regular setup with $M = K/N$. As server $n$ is storing $U_n$ (which is uniformly distributed over $\mathbb{F}_q$), $\forall n \in [N]$ we have $\eta_n = \frac{H(U_n)}{L} = \frac{1}{L}$.

\section{Proof of Optimality of Rate}
\label{converse}

    In this section we prove that the achieved rate $ML/K = L/N$ is optimal for our $(K,N,\frac{K}{N},L)$ bi-regular PID setup. First we will give a bound on size of transmission that servers associated to message $W_D$, i.e.\ servers indexed by $\mathcal{N}_D$, have to do in order to convey $W_D$ correctly (i.e.\ to satisfy \eqref{correctness}).
    \begin{lemma}\label{send_enf_info}
        Message $W_D$ can be correctly conveyed to the user only if 
        \[
            \sum_{n \in \mathcal{N}_D} T_n \geq H(W_D)
        \]
    \end{lemma}
    \begin{prof}
        From \eqref{correctness} we know that 
        \begin{align*}
            &H(W_D | A_{[N]}^D ) = 0 \\
            \implies &I(W_D ; A_{[N]}^D) = H(W_D) \\
            \mbox{Now, }H(W_D) = &I(W_D ; A_{[N]}^D) = I(W_D ; A_{[N] \setminus \mathcal{N}_D}^D ) \\
            &+ I(W_D ; A_{\mathcal{N}_D}^D| A_{[N] \setminus \mathcal{N}_D}^D)
        \end{align*}
        Consider the term $I(W_D ; A_{[N] \setminus \mathcal{N}_D}^D )$. Transmissions $A_{[N] \setminus \mathcal{N}_D}^D$ are only the functions of server storages $Z_{[N] \setminus \mathcal{N}_D}$, index $D$, and shared randomness $U$, which are all independent of message $W_D$ and hence we get 
        \[
            I(W_D ; A_{[N] \setminus \mathcal{N}_D}^D ) \leq I(W_D ; Z_{[N] \setminus \mathcal{N}_D} , U , D) = 0.
        \]
        Substituting this value above, we get
        \begin{align*}
            H(W_D) &= I(W_D ; A_{\mathcal{N}_D}^D| A_{[N] \setminus \mathcal{N}_D}^D) \\
            &= H(A_{\mathcal{N}_D}^D | A_{[N] \setminus \mathcal{N}_D}^D) - H(A_{\mathcal{N}_D}^D | W_D, A_{[N] \setminus \mathcal{N}_D}^D) \\
            &\leq H(A_{\mathcal{N}_D}^D | A_{[N] \setminus \mathcal{N}_D}^D) \leq H(A_{\mathcal{N}_D}^D) \leq \sum_{n \in \mathcal{N}_D} T_n \\
        \end{align*} \hfill \qedsymbol
    \end{prof}

    Now we proceed to show that, in order to hide the index of message $W_D$, the set of servers associated to messages other than $W_D$, i.e.\ servers associated to messages $W_k$ where $k \in [K] \setminus \{D\}$, are also required to perform transmissions of size more than $H(W_k)$. This is formalized in Lemma~\ref{send_confusing_info}
    \begin{lemma}\label{send_confusing_info}
        Message $W_D$ can be conveyed privately to the user only if 
        \[
            \sum_{n \in \mathcal{N}_k} T_n \geq H(W_k) , ~~ \forall k \in [K] \setminus \{D\}
        \]
    \end{lemma}
    \begin{prof}
        Consider for some $k \in [K]$
        \[
            \sum_{n \in \mathcal{N}_k} T_n < H(W_k). 
        \]
        Then, from Lemma~\ref{send_enf_info} we know that message $W_k$ cannot be conveyed correctly if $\sum_{n \in \mathcal{N}_k} T_n < H(W_k)$. So, user can infer that message $W_k$ is not being conveyed to it and that $D \in [K] \setminus \{k\}$, violating privacy constraint. \hfill \qedsymbol
    \end{prof}
    Lemma~\ref{send_enf_info} give a lower bound on size of transmissions that have to be done by the servers indexed by $\mathcal{N}_D$ and Lemma~\ref{send_confusing_info} gives a lower bound on the size of transmissions that have to be done by the servers indexed by $\mathcal{N}_k$ for $k \in [K] \setminus \{D\}$ in order to convey message $W_D$ to the user correctly and privately.

    Now we prove Theorem \ref{capacity}.
    \begin{prof}
        We know from Lemma~\ref{send_enf_info} and Lemma~\ref{send_confusing_info} that $W_D$ can be conveyed to the user correctly and privately only if
        \[
            \sum_{n \in \mathcal{N}_k} T_n \geq H(W_k) , ~~~ \forall k \in [K]
        \]
        summing both sides over all $k \in [K]$ we get 
        \[
            \sum_{k \in [K]} H(W_k) = KL \leq \sum_{k \in [K]} \sum_{n \in \mathcal{N}_k} T_n.
        \]
        Considering the term on the RHS, as each server is associated to $ML$ messages, the term corresponding to transmission of the $n^{th}$ server i.e.\ $T_n$ will appear $ML$ times in the summation. Therefore we have
        \begin{align*}
            KL \leq &\sum_{n \in [N]} ML T_n = ML \sum_{n \in [N]} T_n \\
            \implies &KL \leq ML \sum_{n \in [N]} T_n \\ 
            \implies &\frac{L}{\sum_{n \in [N]} T_n} \leq \frac{ML}{K} \\
            \implies &R \leq \frac{ML}{K} = \frac{L}{N}.  
        \end{align*}
        Since our scheme achieves this rate in the setting considered, we conclude that 
        \[
            C_{CS} = \frac{L}{N} = \frac{ML}{K}    
        \] \hfill \qedsymbol
    \end{prof}

    \textbf{Remark:} Lemma~\ref{send_enf_info} only deals with the correctness of the scheme. If only correctness is required then rate $R = 1$ can be easily achieved. For instance, divide each message into $L$ equal and non overlapping sub-messages, $W_k = \{W_{k,l} : l \in [L]\} ~\forall k \in [K]$. Then store sub-messages of message $W_k$ across the servers indexed by $\mathcal{N}_k$ for all $k \in [K]$. Then, in order to correctly convey message $W_D$, servers indexed by $\mathcal{N}_D$ will transmit the sub-messages of $W_D$, and serves indexed by $[N] \setminus \{D\}$ will not transmit anything. This scheme is correct and achieve rate $R=1$ but the scheme is not private, as the user can infer that only the servers associated to message $W_D$ are transmitting and therefore message $W_D$ is being conveyed. Lemma~\ref{send_confusing_info} deals with privacy, and impose necessary condition on the transmissions on the servers not associated to the message being delivered.

\section*{Acknowledgement}
    This work was supported partly by the Science and Engineering Research Board (SERB) of Department of Science and Technology (DST), Government of India, through J.C. Bose National Fellowship to B. Sundar Rajan, and by the Ministry of Human Resource Development (MHRD), Government of India, through Prime Minister’s Research Fellowship (PMRF) to Kanishak Vaidya.
    
    \bibliographystyle{IEEEtran}
    
\end{document}